\documentclass[aps,prb,twocolumn,floatfix,longbibliography]{revtex4-1}
\usepackage{amsmath,amssymb,graphicx,bm,color}
\def\veck{\mathbf k}
\def\vecq{\mathbf q}

\def\vecQ{\mathbf Q}

\def\be{\begin{equation}}
\def\ee{\end{equation}}

\definecolor{darkred}{rgb}{0.6,0.0,0.0}

\begin{document}
\title{Quantum dynamics in symmetry-breaking states of correlated electrons: Antiferromagnetic phase} 

\author{V\'aclav  Jani\v{s}}  
\email{janis@fzu.cz}

\affiliation{Institute of Physics, The Czech Academy of Sciences, Na Slovance 2, CZ-18200 Praha  8,  Czech Republic}
 
 \author{Mukesh Khanore}

\affiliation{Institute of Physics, The Czech Academy of Sciences, Na Slovance 2, CZ-18200 Praha  8,  Czech Republic}

 \author{Antonín Klíč}

\affiliation{Institute of Physics, The Czech Academy of Sciences, Na Slovance 2, CZ-18200 Praha  8,  Czech Republic}

\date{\today}

%\maketitle

\begin{abstract}
Symmetry-breaking phases in many-fermion systems are characterized by anomalous functions that represent transient processes during which some properties of free particles, such as spin or charge, are not conserved. Connecting the low-temperature symmetry-breaking phase with the high-temperature one within the Baym-Kadanoff scheme, beyond the static mean-field approximation, remains an unresolved, long-standing challenge. We identify the reason why approximations with critical dynamical fluctuations in the Schwinger-Dyson equation lead to a mismatch in the transition temperatures calculated from the high- and low-temperature phases. We propose a solution to this generic problem by excluding anomalous contributions to response functions that do not obey conservation of excitations in their interactions. We illustrate this behavior using the example of an antiferromagnetic state. We reveal that the spectral function in the antiferromagnetic phase exhibits a double-gap structure at zero temperature when the anomalous self-energy is frequency-dependent.           
\end{abstract}
%\pacs{72.15.Qm, 75.20.Hr}

\maketitle %newpage
\section{Introduction} 
Coherence in extended many-particle systems leads to unique situations where quantum dynamics can significantly impact thermodynamic behavior. The interplay between microscopic dynamics and the global macroscopic behavior of many-body systems becomes particularly relevant in the critical regions of transitions to new equilibrium states with broken symmetries. Quantum dynamics is directly responsible for two paradigms of long-range order in correlated electrons: magnetism and superconductivity of metals.      

Most existing approaches to quantum many-body systems treat microscopic dynamics separately from the thermodynamic long-range order. While numerous, more or less advanced schemes have been developed to approximate quantum dynamics in the normal phase, a static mean field remains the only consistent thermodynamic description of the order parameter in symmetry-breaking phases. The static order may be sufficient in weakly correlated systems, but it fails in intermediate and strong coupling. Deviations of the magnetic susceptibility from the Pauli paramagnetic response and the failure of the Stoner criterion indicate that quantum dynamics may non-negligibly affect the global magnetic behavior. It is then inconsistent to combine the Hartree-Fock theory of the ordered phase with strong quantum fluctuations of the high-temperature state since their critical points are significantly distinct.

The first attempts to include quantum dynamics beyond the Stoner model of metallic magnetism were made with the renormalized random-phase approximation \cite{Suhl:1967aa,Levine:1968aa,Hamann:1969aa}. This approximation includes dynamical spin fluctuations but fails to describe the Curie-Weiss susceptibility of transition metals. Moriya, with collaborators, developed a self-consistent renormalized theory that provides a qualitatively correct description of itinerant magnetism consistent with experimental observations of weak ferromagnets \cite{Moriya:1973aa,Moriya:1973ab,Hasegawa:1974aa,Moriya:1985aa}. The next level of dynamical approximations was provided by the dynamical mean-field theory (DMFT) \cite{Georges:1996aa,Kotliar:2001aa,Lichtenstein:2001aa}. Further on, it was demonstrated that renormalized vertex functions with a two-particle self-consistency must be used to reach a more accurate description of strongly correlated electron systems  \cite{Bickers:1989ab,Janis:1999aa,Janis:2007aa,Hirschmeier:2015aa,Rohringer:2018aa,Janis:2020ab,Re:2021aa,Witt:2021aa,Klett:2022aa}.  

All the approaches to implementing quantum dynamics into the many-body response functions have been developed in the high-temperature paramagnetic phase. Extensions of the dynamical approximations with a frequency-dependent self-energy, determined from the Schwinger-Dyson equation containing a singular vertex, encounter inherent problems in the symmetry-breaking phases. The broken symmetry generates new anomalous contributions to the thermodynamic potential, vanishing in the disordered phase \cite{Mattuck:1968aa,Haussmann:1999aa,Janis:2001ab,Janis:2014ab}. Some of the anomalous propagators do not generally vanish at the critical point corresponding to the pole of the two-particle response function of the high-temperature symmetric phase. The mismatch of the critical points from the dynamical vertex of the Schwinger-Dyson equation and the vanishing of the order parameter has been extensively discussed in the search for a consistent dynamical extension of the mean-field theory of superconductivity when studying the BCS-BEC crossover  \cite{Haussmann:1993aa,Micnas:1995aa,Pedersen:1997aa,Keller:1999aa,Chen:2005aa,Lipavsky:2008aa,Levin:2010aa,Kita:2011aa,Sopik:2011aa,He:2016aa,Chen:2024aa}. The existence of two different critical points was suggested to indicate the existence of a first-order transition of the self-consistent T-matrix approximation even though the transition should be continuous \cite{Pieri:2004aa,Haussmann:2007aa,Fukushima:2007aa}.  

The reason for the failure to match the symmetric solution with the one with a broken symmetry at a single transition point is more profound. Quantum many-body models have two different ways to define the two-particle vertex that are compatible with the one-particle self-energy. The first way is to extract the vertex from the microscopic perturbation theory and the dynamical Schwinger-Dyson equation (dynamical vertex). The other indirect way is to use the functional-differential form of the  Ward identity between the self-energy of the perturbation theory and the two-particle irreducible vertex to determine the latter and to make the approximation compatible with macroscopic conservation laws (conserving vertex) \cite{Janis:2024aa}. The two definitions should lead to a unique result in the exact theory. However, we showed earlier that the two ways produce different results in any approximate scheme, including asymptotically exact solutions such as DMFT \cite{Janis:1998aa,Janis:2017aa}. This unavoidable difference leads to ambiguity in determining the critical points indicating the instability of equilibrium phases. The critical points of the dynamical and conserving vertices can be made identical when the Ward identity is applied only to the fluctuations of the symmetry-breaking order parameter, which has odd symmetry with respect to the symmetry-breaking field. The Schwinger-Dyson equation is used only for the normal self-energy with even symmetry  \cite{Janis:2019aa}. This construction makes the conserving and dynamical vertices qualitatively equivalent in the critical region, but only in the high-temperature phase. Another assumption is needed to make the dynamical and conserving vertices qualitatively equivalent in the ordered phase and to ensure that the vanishing of the anomalous propagators (order parameter) is identical to the critical point of the high-temperature susceptibility. 

This paper presents a scheme for consistently extending approximations with the self-energy determined from the Schwinger-Dyson equation with a singular two-particle vertex from the high-temperature phase to the symmetry-breaking phases of correlated electron systems. We analyze the contributions to the thermodynamic potential from the anomalous Green functions within the thermodynamically consistent renormalization scheme of Baym and Kadanoff \cite{Baym:1961aa,Baym:1962aa}. We find that the anomalous terms that do not vanish at the critical point of the high-temperature susceptibility do not guarantee the conservation of the number of electrons and holes in closed systems and must be suppressed to restore continuity at the transition point.  We demonstrate the general scheme on a dynamical mean-field approximation of the Hubbard model and show how many-body quantum dynamics affects the structure of the antiferromagnetic phase. We thereby disclose a previously unnoticed two-gap structure in the zero-temperature spectral function.

\section{Model and approximation with critical quantum dynamics}

\subsection{Model Hamiltonian and thermodynamic potential}

Critical behavior emerges due to long-range fluctuations. It is universal because it does not depend on the precise microscopic structure. To generate many-body quantum critical dynamics, two key components are required: nonlocal kinetic energy and localized interaction, represented by non-commuting operators. The tight-binding Hubbard Hamiltonian characterizes the paradigmatic model of correlated lattice electrons with critical dynamics applicable at any strength of the electron interaction. Its single-band version for spin $s=1/2$ is   
\begin{align}\label{eq:Hubbard}
\widehat{H}_{H}&=\sum_{{\bf k},\sigma} \epsilon({\bf k})
   c^{\dagger}_{{\bf k}\sigma}
  c^{\phantom{\dagger}}_{{\bf k}\sigma}   +
  U\sum_{i}\widehat{n}_{i\uparrow}\widehat{n}_{ i \downarrow}  \,,
\end{align}
where $\epsilon(\veck)$ is the lattice dispersion relation, $ c^{\dagger}_{{\bf k}\sigma}$ and $c^{\phantom{\dagger}}_{{\bf k}\sigma}$ are creation and annihilation operators of electrons with momentum $\veck$  and spin $\sigma$. Further on, $\widehat{n}_{ i\sigma}$ is the operator of the density of particles with spin $\sigma$ in the elementary cell around lattice site $\mathbf{R}_{i}$, and $U$ is the locally averaged electron repulsion due to the screened Coulomb potential.

A consistent description of thermodynamics, including critical behavior and many-body quantum dynamics, must provide a unique definition of all measurable quantities. The scheme of Baym and Kadanoff provides a systematic and controllable method for obtaining thermodynamically consistent approximations from microscopic quantum dynamics. The core of this approach is the grand potential that consists of three parts: the free energy of noninteracting electrons in a dynamical potential, self-energy $\Sigma$, the Legendre term, $G\Sigma$, making the self-energy a variational function together with its Legendre conjugate, one-particle Green function $G$,  and the Luttinger-Ward functional $\Phi$ \cite{Luttinger:1960aa}. The latter is a function only of the interaction and the renormalized Green function in the conserving theories, the so-called $\Phi$-derivable approximations  \cite{Baym:1961aa,Baym:1962aa}.  The full thermodynamic potential can then be represented as      
\begin{multline}\label{eq:PhiG}
 W[G_{\uparrow},\Sigma_{\uparrow},G_{\downarrow},\Sigma_{\downarrow}]  =  - \sum_{\bf k}T\sum_{n=-\infty}^{\infty}\sum_{\sigma = \pm 1}
 \\
   \left\{      
 e^{i\omega_n0^{+}}\ln \left[ i\omega _n+\mu _\sigma -\epsilon  
     ({\bf k}) -\Sigma_\sigma ({\bf
       k},i\omega_n)\right] 
          \right. \\  \left.      \phantom{ e^{i\omega_n0^{+}}}    
       +\ {G}_\sigma ({\bf k}, i\omega_n)\Sigma _\sigma ({\bf k},i\omega_n) \right\} + \Phi [G;U] \,,
\end{multline}
 where $\beta = 1/k_{B}T$, $\mu_{\sigma} = \mu + \sigma h$, $\omega_{n} (2n + 1)\pi T$ are fermionic Matsubara frequencies,  and $h$ is the external magnetic field. We set the universal constants $k_{B}= \hbar = 1$. The equilibrium state is reached when  $\delta W/\delta \Sigma_{\sigma}(\veck,i\omega_{n}) =0$ and $\delta W/\delta G_{\sigma}(\veck,i\omega_{n}) =0$. The former equality is the Dyson equation determining the equilibrium value of $G_{\sigma}(\veck,i\omega_{n})$, and the latter is the dynamical Schwinger-Dyson equation for the self-energy $\Sigma_{\sigma}(\veck,i\omega_{n})$. The Hartree-Fock approximation is obtained when the Luttinger-Ward functional depends only linearly on the interaction strength. The entire impact of the particle interaction on the thermodynamic properties is contained in the Luttinger-Ward functional, the form of which is determined by the chosen approximation derived from the renormalized perturbation theory.

\subsection{FLEX approximation}
 
The simplest approximation with a critical behavior of the Schwinger-Dyson equation is fluctuation exchange (FLEX) \cite{Bickers:1989aa}.  It is an extension of the self-consistent RPA in magnetic systems and the self-consistent $T$-matrix in superconductors, and it is an extension of the weak-coupling Hartree-Fock approximation to intermediate coupling.  The Luttinger-Ward functional for the Hubbard model with repulsive, magnetic interaction, $U>0$, in the spin-symmetric phase, is  
\begin{subequations}\label{eq:phi-FLEX}
 \begin{equation}
\Phi_{FLEX}[G,U]
= \sum_{\vecq}T\sum_{\nu_{m}} \ln\left(1 + U \phi(\vecq,i\nu_{m})\right)\,.
\end{equation} 
with the electron-hole bubble
\begin{multline}
\label{eq:phi-eh-singlet}
\phi(\vecq,i\nu_{m}) 
= \frac 1{ N}\sum_{\veck}T\sum_{\omega_{n}} G(\veck,i\omega_{n})
\\
\times G(\vecQ + \veck + \vecq,i\omega_{n} + i\nu_{m}) \,,
\end{multline}
\end{subequations}
%where $\bar{\sigma} = -\sigma$. 
We denoted the bosonic Matsubara frequencies $\nu_{m}= 2m\pi T$ and extracted momentum $\vecQ$ so that $\phi(\mathbf{0},0)$ is at minimum. 
The Schwinger-Dyson equation for the self-energy  derived from the FLEX generating functional is
\be\label{eq:SDE-even}
\Sigma(\veck,i\omega_{n}) = \frac UN\sum_{\vecq}T\sum_{\nu_{m}}\frac{G(\veck + \vecq,i\omega_{n} + i \nu_{m})}{1 + U\phi(\vecq,i\nu_{m})}\,.
\ee

Fluctuation exchange is a closed-form approximation with an instability of the spin-symmetric, high-temperature state at $1 = U\phi(\mathbf{0},0)$ for a transfer momentum $\vecQ$. It is the critical point of the dynamical vertex in the  Schwinger-Dyson equation~\eqref{eq:SDE-even}. We have the ferromagnetic critical point at $\vecQ=\mathbf{0}$ and the antiferromagnetic instability at $\vecQ = (\pi,\pi,\pi)$. The spin-symmetric solution beyond the critical point of the Schwinger-Dyson equation,  $1 < U\phi(\mathbf{0},0)$, becomes unstable, and a transition to a new state occurs, breaking the spin symmetry. 

Continuing the above FLEX approximation into the ordered phase may seem straightforward. The static description of the ordered phase offered by the Hartree-Fock theory is inadequate. If we proceed standardly, we separate the normal propagator $\bar{G}(\veck,i\omega_{n}) =[G_{\uparrow}(\veck,i\omega_{n}) + G_{\downarrow}(\vecQ + \veck,i\omega_{n})]/2$ with even symmetry with respect to the symmetry-breaking field,  from a dynamical order parameter, a propagator,  $\Delta G(\veck,i\omega_{n}) =[G_{\uparrow}(\veck,i\omega_{n}) - G_{\downarrow}(\vecQ + \veck,i\omega_{n})]/2$ with odd symmetry with respect to the symmetry-breaking field. The electron-hole bubble $\phi$ in Eq.~\eqref{eq:phi-FLEX} is represented via the new even and odd  Green functions, $\bar{G}(\veck, i\omega_{n})$ and $\Delta G(\veck, i\omega_{n})$. The electron-hole bubble in the ordered phase then is 
\begin{multline}
\label{eq:phi-eh-ordered}
\phi(\vecq,i\nu_{m})  \to \phi(\vecq,i\nu_{m}) - \Delta \phi(\vecq,i\nu_{m} 
\\
= \frac 1{ N}\sum_{\veck}T\sum_{\omega_{n}}\left[ \bar{G}(\veck,i\omega_{n}) \bar{G}(\vecQ + \veck + \vecq,i\omega_{n} + i\nu_{m})
\right. \\ \left. 
  -\ \Delta{G}(\veck,i\omega_{n}) 
\Delta{G}(\vecQ + \veck + \vecq,i\omega_{n} + i\nu_{m})\right]
\,.
\end{multline}
The self-energy is also decomposed in the same way into even and odd components, $\Sigma(\veck,i\omega_{n}) =[\Sigma_{\uparrow}(\veck,i\omega_{n}) + \Sigma_{\downarrow}(\vecQ + \veck,i\omega_{n})]/2$, $\Delta\Sigma(\veck,i\omega_{n})=[\Sigma_{\uparrow}(\veck,i\omega_{n}) - \Sigma_{\downarrow}(\vecQ + \veck,i\omega_{n})]/2$ and enter the thermodynamic potential as Legendre conjugates to $\bar{G}(\veck,i\omega_{n})$, $\Delta{G}(\veck,i\omega_{n})$. The generating functional of the ordered phase $ W[\bar{G},\Delta G,\Sigma,\Delta\Sigma ]$ has functions $\bar{G}, \Delta G$, and $\Sigma, \Delta\Sigma$, the variations of which do not change the equilibrium value of the thermodynamic potential. The new Schwinger-Dyson equation for the anomalous self-energy derived from the generating functional in the ordered phase is
\begin{subequations}
\begin{multline}\label{eq:SDE-odd}
\Delta \Sigma(\veck,i\omega_{n}) = - \frac UN\sum_{\vecq}T\sum_{\nu_{m}}
\\
\frac{\Delta G(\veck + \vecq,i\omega_{n} + i \nu_{m})}{1 + U\phi(\vecq,i\nu_{m}) - U\Delta\phi(\vecq,i\nu_{m})}\,.
\end{multline}
The static order parameter in this approximation is 
\be
\Delta = - \frac UN\sum_{\veck}T\sum_{\omega_{n}}\Delta G(\veck,i\omega_{n} ) \,.
\ee
\end{subequations}

The vanishing of the odd propagator and self-energy is not identical with the critical point, $0 = 1 + U_{HT}\phi(\mathbf{0},0)$, in the Schwinger-Dyson equation~\eqref{eq:SDE-even} and of the high-temperature dynamic susceptibility.  The vanishing of the anomalous Green function from the ordered phase is determined from the limit of the order parameter $\Delta$, magnetization, or staggered magnetization,  to zero. It is reached only if the full anomalous self-energy $\Delta\Sigma(\veck,i\omega_{n}) = \Delta + \delta\Sigma(\veck,i\omega_{n})$ derived from the dynamical Schwinger-Dyson equation vanishes. It happens when the average prduct  $TN^{-1}\sum_{\veck,\omega_{n}}\Delta{G}(\veck,i\omega_{n}) \Delta\Sigma(\veck,i\omega_{n}) \le \Delta^{2}\phi(\mathbf{0},0)$ in the limit $\delta\Sigma(\veck,i\omega_{n}) \ll  \Delta$. The anomalous self-energy vanishes at $U>U_{LT}$. The two critical points $U_{HT}$ and $U_{LT}$  are related  in the FLEX approximation as follows
\begin{multline}\label{eq:UHT-ULT}
U_{HT}  \ge \frac{U_{LT}}{|\phi(\mathbf{0},0)|^{2}}\frac{T}{N}\sum_{\veck,\omega_{n}} G(\veck,i\omega_{n})G(\vecQ + \veck,i\omega_{n})
\\
\times \frac{T}{N}\sum_{\veck',\omega_{n'}} \frac{G(\veck',i\omega_{n'})G(\vecQ + \veck',i\omega_{n'})}{1 + U \phi(\veck - \veck^{\prime}, i\omega_{n} - i\omega_{n^{\prime}})} \ge 1\,.
\end{multline}
We see that $U_{HT} \ge U_{LT}$ and they are equal only if the denominator on the right-hand side of the above equation equals one, in the Hartree-Fock approximation. Whenever the self-energy contains dynamical contributions from $U^{2}$ and higher orders, the long-range order emerges at a temperature $T_{L}$, higher than the temperature $T_{H}$ at which the instability of the high-temperature paramagnetic phase occurs. The paramagnetic and magnetic states coexist in the temperature interval $T_{L}< T < T_{H}$, and the transition to the ordered states is of first order with a jump in the magnetization.  This, however, is an unacceptable artifact of the dynamical approximations when continued beyond the instability of the paramagnetic phase in this standard way. 

It is necessary to stress that the discrepancy between the instability of the high-temperature diordered phase at $U_{HT}$ and the instability of the ordered low-temperature phase at $U_{LT}$ is a generic feature of any solution with a critical point in the two-particle vertex. The same reasoning can be used even in the exact solution. There, one replaces $1 + U\phi(\mathbf{0},0)$ with the minimal eigenvalue of the operator in the Hilbert space of fermionic momenta $\veck,\veck'$
\begin{multline}
S_{\vecQ}(\veck,\veck') = \delta_{\veck,\veck'} + T\sum_{\omega_{n}}\Lambda_{\vecQ}(\veck,i\omega_{n},\veck'
,i\omega_{n};\mathbf{0},0)
\\
\times G(\veck',i\omega_{n})G(\vecQ + \veck',i\omega_{n}) \,,
\end{multline}
in the determinantion of $U_{HT}$ from the Schwinger-Dyson equation with $\mathrm{Min}[\hat{S}_{\vecQ}] = 0$ determining the critical point. We denoted $\Lambda_{\vecQ}(\veck,i\omega_{n},\veck'
,i\omega_{n'};\mathbf{q},i\nu_{m})$ the exact electron-hole irreducible vertex. The instability of the ordered phase will be determined from an equation similar to Eq.~\eqref{eq:UHT-ULT}, where the denominator on its right-hand side will be replaced by the Kernel of the Bethe-Salpeter equation for the electron-hole vertex with the irreducible vertex $\Lambda_{\vecQ}(\veck,i\omega_{n},\veck',i\omega_{n'};\mathbf{q},i\nu_{m})$. The inequality $U_{HT} > U_{LT}$ then holds for any solution with a singular Schwinger-Dyson equation for the self-energy. 
%  
%\begin{subequations}
%\begin{multline}
%y(\omega) = - \int_{-\infty}^{\infty}\frac{dx}{\pi}\frac{e^{\beta x}\left(1 + e^{\beta \omega}\right)}{1 + e^{\beta(\omega + x)}}
%\\ 
% \times\Im\left[\frac{1}{\left[x_{+} + \bar{\mu} - \Sigma(x_{+})\right]^{2}  - \epsilon^{2} }\right] 
% \\
%  \times\int_{-\infty}^{\infty}\frac{dy}{\pi}f(-y)\Im G(y_{+})
%  f(y + x) \Im G(y + x_{+})\,.
%\end{multline}
%\end{subequations}

The instability of the ordered phase is identical to the instability of the high-temperature phase only in the linear order of the interaction strength. This means that the static Hartree-Fock theory is the only one that consistently matches the ordered and disordered phases at the critical point. The dynamical approximations containing higher powers of the interaction strength must be appropriately modified in the ordered phase to achieve consistency between the solutions in the ordered and disordered phases. This result applies to any self-energy with a singular Schwinger-Dyson equation. 

It is generally known that there is no unique way to introduce symmetry-breaking anomalous functions into the thermodynamic potential. The standard method of extending the FLEX approximation into the ordered phase, as used above, may be modified by adding other compensating contributions or suppressing some of the existing ones \cite{Janis:2025aa}. We utilize this option and analyze all admissible contributions of the two-particle propagators to the thermodynamic potential with broken symmetry in Appendices A and B, aiming to identify those that vanish exactly at the instability point of the symmetric phase, the critical point of the dynamical vertex. We use physical arguments to justify the proper selection of the symmetry-breaking contributions leading to the continuous matching of the symmetry-breaking phase with the symmetric one at a single transition point.

\section{Ordered phase - Consistent description} 
\subsection{Nambu formalism}

The best way to continue symmetric solutions beyond the Hartree-Fock theory into the ordered phases is to use the Nambu spinor formalism \cite{Nambu:1960aa}. It is a generic scheme covering all types of ordering of systems with two distinguishable species of electrons. We select a spin $\sigma$ as the leading (majority) species and relatively to it we distinguish two normal propagators $G(\veck,i\omega_{n}) = G_{\sigma}(\veck,i\omega_{n})$ and $G_{\vecQ}(\veck,i\omega_{n}) \equiv G_{\bar{\sigma}}(\vecQ + \veck,i\omega_{n})$ that are the diagonal elements of the Nambu spinor.  We denoted $\bar{\sigma} = - \sigma$. The off-diagonal elements of the Nambu spinor are the odd propagators $\Delta G(\veck,i\omega_{n})$. The  self-energies $\Sigma(\veck,i\omega_{n})$,  $\Sigma_{\vecQ}(\veck,i\omega_{n})$, and  $\Delta \Sigma(\veck,i\omega_{n})$ have the same spinor structure.  The thermodynamic potential for the non-interacting part of the ordered state, characterized by the critical momentum $\vecQ$ in the Nambu representation, is  
\begin{widetext}
\begin{multline}\label{eq:W0}
W_{0}[G, G_{\vecQ},\Sigma,\Sigma_{\vecQ},\Delta G,\Delta\Sigma] = \sum_{\veck} T \sum_{\omega_{n}}
\left\{
\mathrm{Tr}\left[ \begin{pmatrix}
G(\veck,i\omega_{n}) \ ,& \Delta G(\veck,i\omega_{n})  \\
\Delta G(\veck,i\omega_{n}) \ , & G_{\vecQ}(\veck,i\omega_{n})
\end{pmatrix}\begin{pmatrix}
\Sigma(\veck,i\omega_{n}) \ ,& \Delta \Sigma(\veck,i\omega_{n})  \\
\Delta \Sigma(\veck,i\omega_{n}) \ , & \Sigma_{\vecQ}(\veck,i\omega_{n})
\end{pmatrix}\right]
\right.\\ \left. 
  -\   e^{i\omega_n0^{+}}\mathrm{Tr}\ln\begin{pmatrix} i\omega_{n} + \mu - \epsilon(\veck) - \Sigma(\veck,i\omega_{n}) \ , &  h + \Delta\Sigma(\veck,i\omega_{n}) \\
h + \Delta\Sigma(\veck,i\omega_{n}) \ , & i\omega_{n} + \mu - \epsilon(\vecQ + \veck) - \Sigma_{\vecQ}(\veck,i\omega_{n}) 
\end{pmatrix}
\right\} \,,
\end{multline}
where $h$ is an auxiliary field generating transitions between the two electron species. Notice that in the case of a ferromagnetic order, $\vecQ=0$, this representation corresponds to a system exposed to a transversal magnetic field. The crucial step in constructing a consistent thermodynamic potential in the ordered phase is the selection of the Luttinger-Ward functional with anomalous contributions that will continuously match the high-temperature solution at the critical point of its instability.  As discussed in Appendix A, one must suppress the anomalous non-conserving contributions in the closed system. A bispinor distinguishing singlet and triplet two-particle propagators, Appendix B, generally represents the full Luttinger-Ward functional in the ordered phase.  It reduces to a  $2\times 2$ spinor form with the following expression when the spin singlet and triplet propagators become indistinguishable, see Appendix B,   
\begin{equation}\label{eq:Phi-matrix}
\Phi_{FLEX}[\phi,\Delta \phi;U] = - \sum_{\vecq}T\sum_{\nu_{m}}\left[U\Delta \phi(\vecq,i\nu_{m})  
%\right. \\ \left.
- \frac 12\mathrm{Tr}\ln  \begin{pmatrix}
1 + U\phi(\vecq,i\nu_{m}) \ ,& U\Delta \phi(\vecq,i\nu_{m})  \\
U\Delta \phi(-\vecq,-i\nu_{m}) \ , &1 + U\phi(-\vecq,-i\nu_{m})\end{pmatrix}\right] \,.
\end{equation}
\end{widetext}

The normal $\phi(\vecq,i\nu_{m})$ and anomalous $\Delta \phi(\vecq,i\nu_{m})$  bubbles are defined as 

\begin{subequations}\label{eq:PhiDeltaphi-eh-gen}
\begin{multline}\label{eq:phi-eh}
\phi(\vecq,i\nu_{m})
 = \frac 1{ N}\sum_{\veck,}T\sum_{\omega_{n}}   G(\veck,i\omega_{n})
 \\
\times G( \vecQ + \veck + \vecq, i\omega_{n} + i\nu_{m}) \,,
\end{multline}

\begin{multline}\label{eq:Deltaphi-eh-gen}
\Delta\phi(\vecq,i\nu_{m})
 = \frac 1{ N}\sum_{\veck,}T\sum_{\omega_{n}}  \Delta G(\veck,i\omega_{n})
 \\
\times\Delta G(\vecQ +  \veck + \vecq, i\omega_{n} + i\nu_{m}) \,.
\end{multline}
\end{subequations}
The Schwinger-Dyson equations for the normal and anomalous self-energies have the following form:
\begin{subequations}
\begin{multline}
\Sigma(\veck,i\omega_{n}) = \frac UN\sum_{\vecq}T\sum_{\nu_{m}}
\\
\frac{G(\veck + \vecq,i\omega_{n} + i \nu_{m})\left(1 + U\phi(\vecq,i\nu_{m})\right)}{\left(1 + U\phi(\vecq,i\nu_{m})\right)^{2} -  U^{2}\Delta\phi(\vecq,i\nu_{m})^{2}} \,,
\end{multline}
\begin{multline}
\Delta \Sigma(\veck,i\omega_{n}) = - \frac UN\sum_{\vecq}T\sum_{\nu_{m}}\left[\Delta G(\veck + \vecq,i\omega_{n} + i\nu_{m} )    \phantom{\frac12}
\right. \\ \left.
+\ \frac{U\Delta\phi(\vecq,i\nu_{m})}{\left(1 + U\phi(\vecq,i\nu_{m})\right)^{2} -  U^{2}\Delta\phi(\vecq,i\nu_{m})^{2}} \right]\,.
\end{multline}
\end{subequations}
The critical change in the Schwinger-Dyson equation in the ordered phase, compared with the standard formulation, is the denominator of the integrand. There is no linear power of the anomalous bubble. This change guarantees that $U_{HT} = U_{LT}$ as we explicitly demonstrate in the following sections. 

The symmetry vector $\vecQ$ strongly affects the ordered phase's structure. Its general form means that one cannot reduce the momentum summations to energy integrals with a density of states or a mean-field approximation. There are, however, two situations, ferromagnetic, $Q=0$, and antiferromagnetic, $\vecQ= (\pi,\pi,\pi)$, that allow for a mean-field-like description with only energy integrals for the bipartite lattices.  We have $\epsilon(\veck) = \epsilon(-\veck)$ for the ferromagnetic phase and  $\epsilon(\vecQ + \veck) = - \epsilon(\veck)$ for the antiferromagnetic order.

\subsection{Antiferromagnetic state: Dynamical mean-field approximation}. 
We apply the general formalism to the antiferromagnetic phase of the Hubbard model within the dynamical mean-field theory. This means that the normal and anomalous self-energies are local and only frequency-dependent. We can explicitly analytically continue all the functions from the Matsubara frequencies to obtain their spectral presentations. The equation for the transition temperature to the antiferromagnetic phase at $h=0$ is $1 + U\phi(\mathbf{0},0)$, which, in the dynamical mean-field (local) approximation, explicitly reads  
\begin{multline}
1 =  U\int_{-\infty}^{\infty}\frac{d\omega}{\pi}f(\omega) \int_{-\infty}^{\infty}d\epsilon \rho(\epsilon)
\\
\times\Im\left[\frac{1}{\left[\omega_{+} + \bar{\mu} - \Sigma(\omega_{+})\right]^{2}  - \epsilon^{2} }\right] \,,
\end{multline}
where $\bar{\mu } = \mu - Un/2 $ is the effective chemical potential, $n$ is the charge density fixing the chemical potential, and $f(x)$ is the Fermi spectral function.     

The local normal and anomalous Green functions in the ordered phase within the DMFT are explicitly      
\begin{subequations}\label{eq:DMFT-GF}
\begin{multline}
G(\omega_{+}) = \int_{-\infty}^{\infty}d\epsilon \rho(\epsilon)
\\
\times\frac{\omega + \bar{\mu}  - \Sigma(\omega_{+})}{\left(\omega_{+} + \bar{\mu} - \Sigma(\omega_{+})\right)^{2}  - \epsilon^{2}  - \left[\Delta + \delta\Sigma(\omega_{+})\right]^{2}}\,,
\end{multline}
\begin{multline}
\Delta G(\omega_{+}) =   \int_{-\infty}^{\infty}d\epsilon \rho(\epsilon)
\\
 \times\frac{\Delta + \delta\Sigma(\omega_{+})}{\left(\omega_{+} + \bar{\mu} - \Sigma(\omega_{+})\right)^{2}  - \epsilon ^{2} -   \left[\Delta + \delta\Sigma(\omega_{+})\right]^{2}}\,,
\end{multline}
\end{subequations}
where $\omega_{+} = \omega + i0^{+}$ denotes the limit to the real axis from the upper complex half-plane. We split the anomalous self-energy into its static part $\Delta$ and a dynamical correction $\delta\Sigma(\omega_{+})$ in the same way we extracted the static Hartree contribution, $Un/2$, from the normal self-energy.  

The Schwinger-Dyson equations for the normal and anomalous self-energies conclude the FLEX approximation in the ordered phase. They are straightforwardly derived from the stationarity point of the generating thermodynamic potential from Eq.~\eqref{eq:W0} and Eq.~\eqref{eq:Phi-matrix}. The normal self-energy in the DMFT is  
\begin{align}\label{eq:SDE-Normal}
\Sigma(\omega_{+}) &=  \int_{-\infty}^{\infty}\frac{dx}{\pi}\left[b(x) G(\omega_{+} + x)  \Im \Gamma(x_{+}) 
\nonumber \right. \\  &\qquad \left.
  -\ f(x) \left[\Gamma(x -  \omega_{+}) - U\right]\Im G( x_{+}) \right]\,,
\end{align}
where $\Gamma(i\nu_{m}) = N^{-1}\sum_{\vecq}\delta \Phi/\delta \phi(\vecq,i\nu_{m})$ is the normal local two-particle vertex and $b(x)$ is the Bose spectral function.. After analytic continuation, the mean-field two-particle vertex reads 
\begin{multline}
\Gamma(\omega_{+}) 
\\
= \frac 1N\sum_{\vecq} \frac{U\left[1 + U\phi(\vecq,\omega_+)\right]}{\left[1 + U\phi(\vecq,\omega_+)\right]^2 - U^2\Delta\phi(\vecq,\omega_+)^2} \,.
\end{multline}

The static anomalous self-energy, macroscopic thermodynamic order parameter $\Delta$ and its microscopic dynamical correction $\delta \Sigma(\omega_{+})$ are determined, after analytic continuation,  from the following equations 
\begin{subequations}\label{eq:SDE-Anomalous}
\begin{align}
\Delta & = U  \int_{-\infty}^{\infty}\frac{dx}{\pi}f(x) \Im\Delta G( x_{+}) \,,
\\
\delta \Sigma(\omega_{+}) &= \int_{-\infty}^{\infty}\frac{dx}{\pi}\left\{b(x)\Delta G(\omega_{+} + x)  \Im \Delta\Gamma(x_{+}) 
\nonumber \right. \\  & \left.  
  -\ f(x) \left[\Delta\Gamma(x -  \omega_{+}) - U\right] \Im\Delta G( x_{+}) \right\}\,.
\end{align}
\end{subequations}
Analogously,  $\Delta\Gamma(i\nu_{m}) = N^{-1}\sum_{\vecq}\delta \Phi/\delta \Delta\phi(\vecq,i\nu_{m})$ is the mean-field anomalous vertex. Its explicit form after analytic continuation is
\begin{multline}
\Delta\Gamma(\omega_+)
\\
 =  - \frac 1N\sum_{\bf q} \frac{U^2 \Delta\phi({\bf q},\omega_+)}{\left[1 + U\phi({\bf q},\omega_+)\right]^2 - U^2\Delta\phi({\bf q},\omega_+)^2} \,.
\end{multline}
We see that the anomalous two-particle vertex, determining the microscopic dynamical correction to the static order parameter, is proportional to the second power of the interaction strength. Hence, it is missing in the Hartree-Fock mean-field approximation.  We apply the Schwinger-Dyson equations only to the imaginary part of the self-energies. The respective real parts are calculated from the Kramers-Kronig relations to guarantee self-energies' analyticity in numerically approximated integrals.

\section{Numerical results}

We solve the Schwinger-Dyson equations for the normal self-energy $\Sigma(\omega_{+})$,  Eq.~\eqref{eq:SDE-Normal},  for the order parameter $\Delta$, and the anomalous dynamical self-energy $\delta\Sigma(\omega_{+})$, Eq.~\eqref{eq:SDE-Anomalous}, in the antiferromagnetic phase of the half-filled Hubbard model, $\bar{\mu}=0$, at a moderate interaction strength $U= 2w$, where $w$ is the energy band halfwidth. We resort to the second-order DMFT with two-particle vertices $\Gamma(\omega_{+}) = U -   U^{2}\phi(\omega_{+})$ and $\Delta\Gamma(\omega_{+})= - U^{2}\Delta\phi(\omega_{+})$. This approximation delivers, in intermediate coupling, qualitatively the same results as the full FLEX approximation within the DMFT. It represents the exact weak-coupling limit of the full DMFT solution. We utilize the semi-elliptic density of states in our numerical calculations.  

\begin{figure}
\includegraphics[width=80mm]{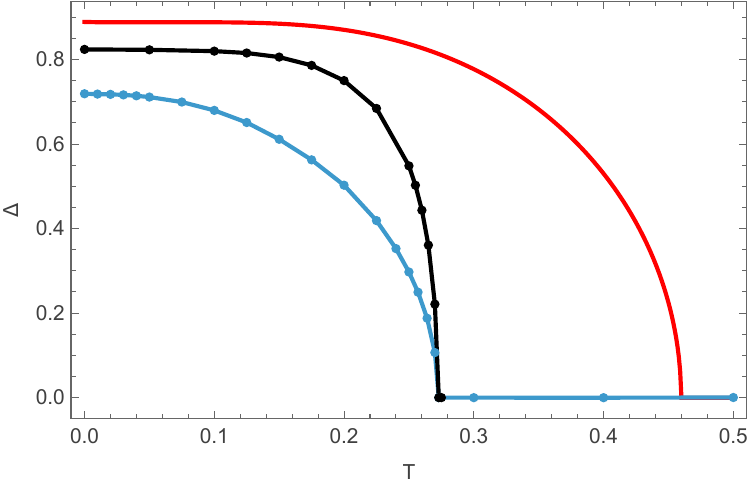}
\caption{Antiferromagnetic order parameter $\Delta$ of the half-filled Hubbard model at $U=2w$, of the second-order approximation with $\Im\Sigma(\omega)<0$, obtained for $\delta\Sigma(\omega)=0$, lower, blue, curve with dots,  and the full dynamical solution $\Delta\Sigma(\omega)$, upper, black, curve with dots, compared with the Hatree-Fock solution, red curve.  The temperature is normalized to the half bandwidth, $w=1$.\label{fig:PhDiag}}
\end{figure}

We plotted in Fig.~\ref{fig:PhDiag} the order parameter $\Delta$. We calculated this parameter in two versions of the solution in the antiferromagnetic phase. One is the solution where the contribution from the dynamical part of the anomalous self-energy $\delta\Sigma(\omega)$ is neglected, blue curve. The complete solution, black curve, with the dynamical anomalous self-energy shares the critical asymptotics at the transition point but amplifies the order parameter at low temperatures. The dynamical solution delivers significantly lower transition temperature than the static Hartree-Fock approximation, red curve. The slope of the order parameter in the dynamical approximation is then significantly steeper below the transition point than in the static solution. Notice that we incorrectly scaled the critical point of the dynamical solution in our earlier paper \cite{Janis:2025aa}.  All solutions deliver the unique critical point, confirming that the transition to the magnetically ordered phase is continuous.   

\begin{figure}
\includegraphics[width=80mm]{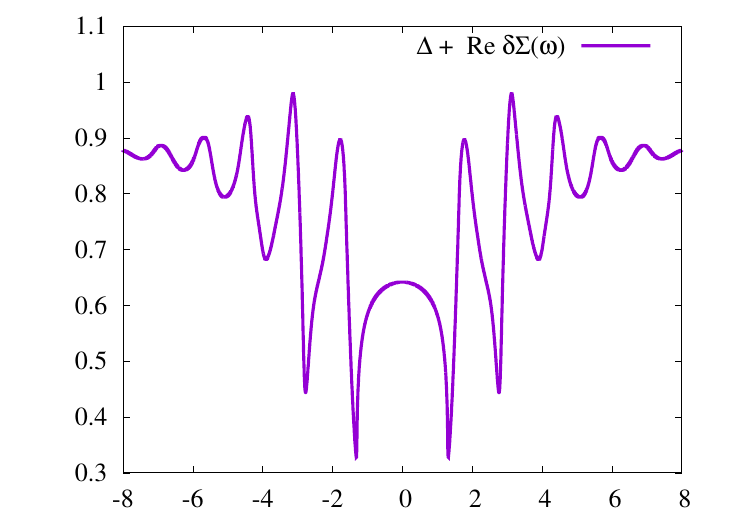}
\caption{Dynamical order parameter $\Delta + \Delta \Re\Sigma(\omega)$ as function of frequency at zero tempearture and $U= 2w$. The energy scale was set $w=1$. \label{fig:D+DX}}
\end{figure}

The dynamics of the anomalous self-energy makes the order parameter frequency-dependent. The dependence of the dynamical order parameter $\Delta +  \Re\delta\Sigma(\omega)$ at low temperatures displays an oscillating structure with a sharp dip from the central maximum towards the edge of the dynamical gap with $\Im\Sigma(\omega) = 0$, Fig.~\ref{fig:D+DX}.  

%a surprisingly non-trivial temperature dependence. 

%The principal difference between the static and the dynamical solution is that $\Delta$ of the dynamical solution is not the gap's halfwidth at zero temperature. The actual gap around the Fermi energy is restricted to . 
%

Quantum fluctuations generate the same dynamical gap in the imaginary part of the normal, $Y(\omega) = - \Im \Sigma(\omega_{+})$ and anomalous, $\Delta Y(\omega) = \Im \delta\Sigma(\omega_{+})$ self-energies.
\begin{figure}
\includegraphics[width=75mm]{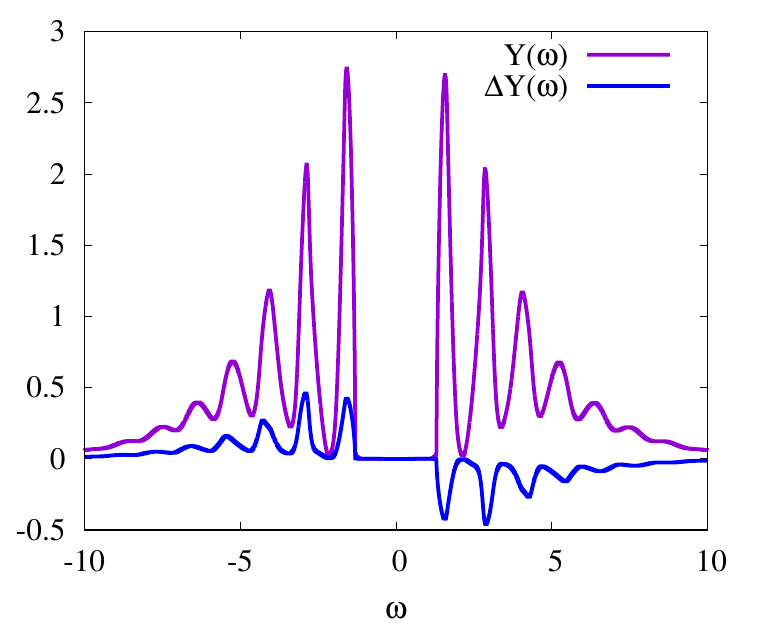}
\caption{The imaginary part of the normal self-energy $Y(\omega) = - \Im \Sigma(\omega_{+})$ and of the anomalous self-energy $\Delta Y(\omega) = \Im\Delta\Sigma(\omega_{+})$ at zero tempearture and for $U=2w$ with the energy scale $w=1$. \label{fig:Y-DY}}
\end{figure}
The normal self-energy contains even powers of $\Delta$ while the anomalous one odd powers, starting with $\Delta^{3}$. The self-energies are convolutions of the imaginary parts of the Green function, and consequently, they have a bigger gap than the spectral function. More importantly, they oscillate, and their amplitude decreases as $\omega^{-2}$ as $|\omega|\to\infty$, see Fig.~\ref{fig:Y-DY}.  

The dynamical gap contains bands of in-gap states with the inner-edge square root singularity for $\left[\omega - \Re\Sigma(\omega\right]^{2} - \left[\Delta + \Re\delta \Sigma(\omega)\right]^{2} \in (0,w^{2})$. The in-gap states are of the Hartree-Fock form since $Y(\omega)=0$, meaning that there are no quantum fluctuations within the dynamical gap. Although the lowest lying excited states are of Hartree-Fock character, the gap separating these states from the Fermi energy is much smaller than the critical temperature, $\omega_0\approx 0.45T_c$ in the second-order approximation. This deviation from the mean-field prediction can be used as one of the hallmarks indicating a non-negligible impact of strong electron correlations. 

\begin{figure}
\includegraphics[width=80mm]{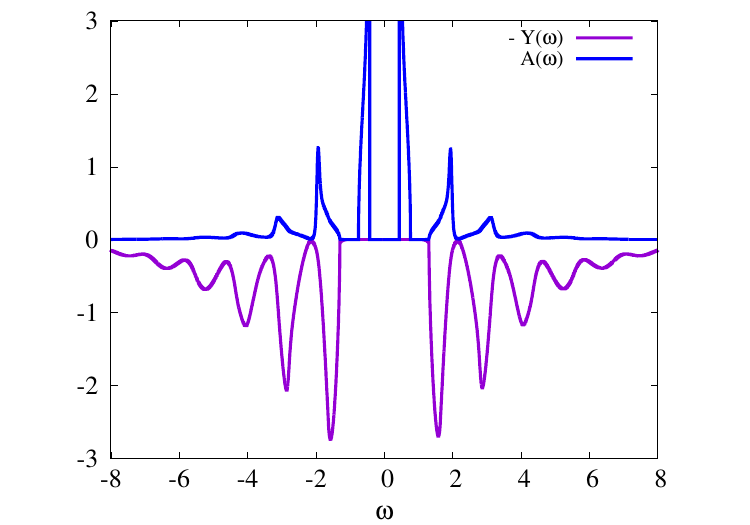}  
\caption{Frequency dependence of the zero-temperature imaginary parts of the normal self-energy $-Y(\omega)$  and the spectral function $A(\omega)$. The band of occupied states within the self-energy gap is Hartree-Fock-like, with no quantum dynamics.     \label{fig:Y-ImG}}
\end{figure}

The spectral function $A(\omega) = - \Im G(\omega_{+})/\pi$ together with the imaginary part of the self-energy $Y(\omega)$ are plotted in Fig.~\ref{fig:Y-ImG}. The dynamical gap is triple the spectral gap in the second-order approximation.  Corrections to the Hartree-Fock spectral function lead to new structures only outside the gap of the self-energy where $Y(\omega)>0$. The existence of a large dynamical gap of the self-energy with in-gap Hartree-Fock-like states is generic whenever quantum dynamics is introduced in the antiferromagnetic phase. It means, whenever the self-energy is frequency-dependent.       

%The normal and anomalous spectral functions at zero temperature are plotted in Fig.~\ref{fig:ImG-ImDG}. We can see two separated bands of occupied states. The band closer to the Fermi energy has a Hartree-Fock structure with square-root singularity since it lies within the gap of the self-energy.  The interaction drives the electrons from the Fermi energy and generates a separated band structure outside the gap of the self-energy.
%
%\begin{figure}
%\includegraphics[width=70mm]{Im_G_dG} 
%\caption{Frequency dependence of the normal and anomalous spectral functions at zero temperature. The spectral function has a two-gap structure with a Hartree-Fock band of occupied states within the larger gap of the self-energy.      \label{fig:ImG-ImDG}}
%\end{figure}

%\begin{figure}
%\includegraphics[width=70mm]{y_delta_y}
%\caption{  \label{fig:PhDiag}}
%\end{figure}

\section{Conclusions}

There is no single method beyond static mean-field theory to extend microscopic quantum dynamics from the high-temperature phase to the ordered phase, characterized by symmetry-breaking long-range thermodynamic order. The static approximation cannot, however, be used if strong electron correlations significantly affect the magnetic response already at high temperatures, resulting in the Curie-Weiss susceptibility. The Baym-Kadanoff scheme provides a consistent and controlled method for incorporating electron correlations into thermodynamic functions. It is, however, consistent and unambiguous only in the high-temperature phase with no critical instability. The application of a symmetry-breaking field can cause severe problems and lead to unphysical behavior if not properly treated. If we include all admissible contributions, Feynman diagrams generated by the symmetry-breaking field, we find a temperature interval in which the ordered and disordered states coexist. The instability of the high-temperature phase occurs below the temperature at which thermodynamic order disappears. This would indicate a first-order, discontinuous transition, which is unphysical.  This behavior is symptomatic of any, even exact, solution with a dynamical self-energy determined from a Schwinger-Dyson equation with a critical point. To achieve consistency between the ordered and disordered phases, the anomalous contributions that break the symmetry of the high-temperature phase must be selected carefully so that all multiparticle processes take place only in the appropriate representation space and the number of the excited states is conserved. Some anomalous contributions must be suppressed, even if they are of the same order as others.    

In this paper, we presented a general construction of consistent thermodynamic theories with long-range order and nontrivial quantum dynamics. We identified the anomalous two-particle propagators contributing to the thermodynamic potential that do not guarantee the two interacting (virtual) particles are from the same representation space of a closed system. These propagators are characterized by the fact that the two-particle functions consist of odd powers of the anomalous Green functions. They must be suppressed to align the ordered and disordered phases at the single transition point. We achieve this if the two-particle Nambu spinor formalism replaces the single-particle description of the symmetric phase with normal and anomalous propagators in the ordered phase. This approach applies to all symmetry-breaking states, whether the ordered phase is reached spontaneously or through the application of an external force.

We applied the general construction to the antiferromagnetic phase of the Hubbard model. We resorted to the FLEX-DMFT as the simplest approximation with a singular dynamical vertex and a local dynamical self-energy. We found that the antiferromagnetic phase of the dynamical approximations has, generically, a two-gap structure. The dynamical gap with $\Im\Sigma(\omega)=0$ is bigger than the spectral gap for $\Im G(\omega)=0$. A band of occupied states emerges within the dynamically induced gap in the self-energy. The in-gap states form a renormalized Hartree-Fock band with no quantum dynamical corrections, $\Im\Sigma(\omega)=0$ within the band. Quantum fluctuations affect only the high-energy excitations lying beyond the dynamical gap.  This feature is universal and holds for any thermodynamically consistent theory. Consequently, the spectral gap is much smaller than the static order parameter $\Delta$, the energy gap of the static Hartree-Fock approximation,  or the critical temperature $T_{c}$, even if we neglect the dynamical anomalous self-energy $\delta\Sigma(\omega)$. The latter is proportional to $\Delta^{3}$.  The two-gap structure with in-gap bands will also appear in the superconducting phase, where quantum dynamical fluctuations become relevant beyond the static BCS mean-field theory.           

\section*{Acknowledgment}

VJ thanks the Cost Action CA 21144, grant INTER-COST LUC24139 of the Czech Ministry of Education, Youth, and Sport for financial support. 

%\begin{widetext}

\appendix
\section{Propagators not conserving spin: Spinor representation}
Phases with broken symmetry distinguish propagators of two species of electrons.  Transitions between them do not conserve spin in magnetism or charge in superconductors. They are spin-flip processes in magnetically ordered states, and they annihilate and create Cooper pairs in superconductivity.  The universal description of states with two species of fermions offers the Nambu spinor formalism \cite{Nambu:1960aa}. It introduces $2\times 2$ spinors for the one-electron propagators containing both spins to allow spin-flip processes in the magnetic systems. The matrix elements of the particle Nambu spinor are defined  in the space-time basis as follows
\begin{equation}
G_{\sigma\sigma^{\prime}}(1 - 2) = - \left\langle \mathrm{T}\left[c_{\sigma}^{\phantom{\dagger}}(1)c_{\sigma^{\prime}}^{\dagger}(2)\right] \right\rangle \,,
\end{equation}
where $1,2$ stand for $\mathbf{R}_{1,2}, \tau_{1,2}$ with imaginary time  $\tau \in (0,\beta)$. Symbol $\mathrm T$ denotes time ordering and $c_{\sigma}^{\phantom{\dagger}},c_{\sigma}^{\dagger}$ are the annihilation and creation operators of the electron with spin $\sigma$. The angular brackets stand for the averaging over the thermal fluctuations. 

The one-particle functions are not enough to describe the response to external excitations.  We also need holes.  The Nambu spinor of a hole analogously is 
\begin{multline}
\bar{G}_{\sigma\sigma^{\prime}}(1 - 2) = - \left\langle \mathrm{T}\left[d_{\sigma}^{\phantom{\dagger}}(1)d_{\sigma^{\prime}}^{\dagger}(2)\right] \right\rangle
\\
 = - G_{\sigma^{\prime}\sigma}(2 - 1) \,,
\end{multline}
with $d_{\sigma}^{\phantom{\dagger}},d_{\sigma}^{\dagger}$  creation and annihilation operators of a hole. The creation of a hole is the same as the annihilation of an electron and vice versa. The last equation follows from the electron-hole symmetry. 

The Nambu spinors of the particle and hole propagators can be represented diagrammatically. Using the above definitions, we have the following graphical representation of a particle
\begin{widetext}
\begin{subequations}
\begin{equation}
\begin{minipage}{30mm}
$$
\begin{pmatrix}
- \left\langle \mathrm{T}\left[c_{\sigma}^{\phantom{\dagger}}(1)c_{\sigma}^{\dagger}(2)\right] \right\rangle \ ,& - \left\langle \mathrm{T}\left[c_{\bar{\sigma}}^{\phantom{\dagger}}(1)c_{\sigma}^{\dagger}(2)\right] \right\rangle 
\\[2pt]
- \left\langle \mathrm{T}\left[c_{\sigma}^{\phantom{\dagger}}(1)c_{\bar{\sigma}}^{\dagger}(2)\right] \right\rangle \ ,& - \left\langle \mathrm{T}\left[c_{\bar{\sigma}}^{\phantom{\dagger}}(1)c_{\bar{\sigma}}^{\dagger}(2)\right] \right\rangle 
\end{pmatrix}  = \begin{pmatrix}
G_{\sigma\sigma}(1 - 2) \ ,&  G_{\bar{\sigma}\sigma}(1 - 2)
\\
G_{\sigma\bar{\sigma}}(1 - 2)\ ,& G_{\bar{\sigma}\bar{\sigma}}(1 - 2)
\end{pmatrix}
\quad = \quad
$$
\end{minipage}\hspace{100mm}
\begin{minipage}{35mm}
\includegraphics[width=35mm]{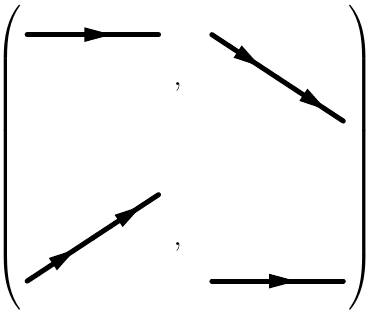}
\end{minipage}
\end{equation}
\\
and of a hole 
\begin{equation}
\begin{minipage}{30mm}
$$
\begin{pmatrix}
\left\langle \mathrm{T}\left[c_{\bar{\sigma}}^{\phantom{\dagger}}(2)c_{\bar{\sigma}}^{\dagger}(1)\right] \right\rangle \ ,&  \left\langle \mathrm{T}\left[c_{{\sigma}}^{\phantom{\dagger}}(2)c_{\bar{\sigma}}^{\dagger}(1)\right] \right\rangle 
\\[2pt]
 \left\langle \mathrm{T}\left[c_{\bar{\sigma}}^{\phantom{\dagger}}(2)c_{{\sigma}}^{\dagger}(2)\right] \right\rangle \ ,&  \left\langle \mathrm{T}\left[c_{{\sigma}}^{\phantom{\dagger}}(1)c_{{\sigma}}^{\dagger}(1)\right] \right\rangle 
\end{pmatrix} = \begin{pmatrix}
\bar{G}_{\bar{\sigma}\bar{\sigma}}(1 - 2) \ ,&  \bar{G}_{\bar{\sigma}\sigma}(1 - 2)
\\
\bar{G}_{\sigma\bar{\sigma}}(1 - 2)\ ,& \bar{G}_{{\sigma}{\sigma}}(1 - 2)
\end{pmatrix}
\quad = \quad
$$
\end{minipage}\hspace{100mm}
\begin{minipage}{35mm}
\includegraphics[width=35mm]{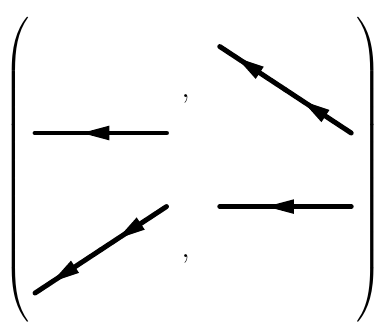}
\end{minipage}
\end{equation}
\end{subequations}
\end{widetext}
We denoted $\bar{\sigma} = - \sigma$. Each matrix element has four points positioned left and right in the upper and lower rows. The right points have coordinate $1$ and the left $2$. The upper row is used for the coordinates of the leading (majority) spins, while the lower row is used for the minority spins. Just two points are connected with an oriented line of a one-particle propagator. The arrows indicate charge propagation with time developing from left to right.  The arrows pointing from right to left denote the propagation of a hole.  The double arrow of the skew lines indicates a spin flip, changing the spin by two. The arrows pointing from left to right indicate changing the spin by removing a spin from the minority spins and adding it to the majority spins. The downward arrows do the opposite. 

%\quad ,\quad \includegraphics[width=40mm]{FLEX-Mag-H}
%\caption{Graphic representation of charge propagation indicated by arrows in the spinor form, left panel. Spins $\uparrow$ and $\downarrow$ are represented by upper and lower lines, respectively. The skew lines stand for spin-flip processes.  The hole propagation is in the right panel.  \label{fig:FEynm1P}}
%\end{figure}
 
The diagonal elements are normal spin-conserving propagators and can be characterized by a single spin index.  The off-diagonal elements are anomalous propagators connecting opposite spins in the Nambu spinors. They are dynamical extensions of the static order parameter and are nonzero only in the ordered phase. We generally need two spin indices to specify the spin-flip processes. We can return to a single-spin denotation of the anomalous spin-dependent particle and hole propagators, the off-diagonal elements of the Nambu spinor,    
\begin{subequations}the\begin{align}
\Delta G_{\sigma}(1 - 2) &= G_{\sigma\bar{\sigma}}(1 - 2) \,,
\\
\Delta \bar{G}_{\sigma}(1 - 2) &= - G_{\bar{\sigma}\sigma}(2 - 1) \,.
\end{align}
\end{subequations}
 We again used the electron-hole symmetry.

\section{Spin-denendent electron-hole propagators -- Bispinor representation}.

Two-particle propagators are diagrammatically composed of two lines. Electron-hole propagators consist of one electron and one hole line with arrows pointed in opposite directions. We must combine normal and anomalous propagators beyond the normal ones in the low-temperature ordered phase. The ordered phase generates different types of anomalous two-particle propagators.  

The spin singlet and triplet normal electron-hole propagators in the momentum-frequency representation are
\begin{subequations}
\begin{align}
\label{eq:phi-eh-s}
{\phi}^{s}_{\sigma}(\vecq,i\nu_{m}) 
&= \frac 1{ N}\sum_{\veck}T\sum_{\omega_{n}} G_{\sigma}(\veck,i\omega_{n})
\nonumber \\ &\qquad
\times G_{\bar{\sigma}}(\vecQ + \veck + \vecq,i\omega_{n} + i\nu_{m}) \,. 
\\ \label{eq:phi-eh-t}
{\phi}^{t}_{\sigma}(\vecq,i\nu_{m}) 
&= \frac 1{ N}\sum_{\veck}T\sum_{\omega_{n}}G_{\sigma}(\veck,i\omega_{n}) 
\nonumber \\ &\quad
\times G_{{\sigma}}(\vecQ + \veck + \vecq,i\omega_{n} + i\nu_{m}) \,
\end{align}
\end{subequations}

Anomalous electron-hole propagators contain combinations of a spin-dependent anomalous propagator with either a normal or another anomalous propagator. The electron-hole propagator containing one normal and one anomalous propagator in the spin singlet and triplet form is
\begin{subequations}
\begin{align}
\label{eq:bar-psi-eh}
\Delta{\psi}^{s}_{\sigma}(\vecq,i\nu_{m}) 
&= \frac 1{ N}\sum_{\veck}T\sum_{\omega_{n}} G_{\sigma}(\veck,i\omega_{n})
\nonumber \\ &\quad
\times
\Delta G_{\bar{\sigma}}(\vecQ + \veck + \vecq,i\omega_{n} + i\nu_{m}) \,, 
\end{align}
\begin{align}
\label{eq:psi-eh}
\Delta\psi^{t}_{\sigma}(\vecq,i\nu_{m}) 
&= \frac 1{ N}\sum_{\veck}T\sum_{\omega_{n}} G_{\sigma}(\veck,i\omega_{n})
\nonumber \\ &\quad
\times
\Delta G_{\sigma}(\vecQ + \veck + \vecq,i\omega_{n} + i\nu_{m}) \,.
\end{align}
\end{subequations}

Finally, we have singlet and triplet electron-hole propagators combining pairs of the anomalous propagators 
\begin{subequations}
\begin{align}\label{eq:bar-phi-eh}
\Delta {\phi}^{s}_{\sigma}(\vecq,i\nu_{m}) 
&= \frac 1{ N}\sum_{\veck}T\sum_{\omega_{n}} \Delta G_{\sigma}(\veck,i\omega_{n})
\nonumber \\ &\quad
\times
\Delta G_{\bar{\sigma}}(\vecQ + \veck + \vecq,i\omega_{n} + i\nu_{m}) \,,
\\
\label{eq:Deltaphi-eh}
\Delta \phi^{t}_{\sigma}(\vecq,i\nu_{m}) 
&= \frac 1{ N}\sum_{\veck}T\sum_{\omega_{n}} \Delta G_{\sigma}(\veck,i\omega_{n})
\nonumber \\ &\quad
\times
\Delta G_{\sigma}(\vecQ + \veck + \vecq,i\omega_{n} + i\nu_{m}) \,.
\end{align}
\end{subequations}

We generally must keep the spin dependence of the anomalous propagators to cover systems with a spin bias. There is no difference between the spin singlet and spin triplet two-particle propagators in the spin-symmetric situations. We can then neglect the spin index at the two-particle propagators. The spinor of the electron-hole propagator standardly used in the ordered phase, Eq.~\eqref{eq:phi-eh-ordered}, is
\begin{multline}\label{eq:phi-Deltapsi}
\begin{minipage}{30mm}
$$
\begin{pmatrix}
\left\langle G_{{\sigma}}\bar{G}_{\bar{\sigma}} \right\rangle \ ,&  \left\langle G_{{\sigma}}\Delta\bar{G}_{\bar{\sigma}}\right\rangle 
\\[2pt]
 \left\langle \bar{G}_{\bar{\sigma}}\Delta{G}_{{\sigma}} \right\rangle \ ,&  \left\langle  \Delta G_{{\sigma}}\Delta\bar{G}_{\bar{\sigma}}\right\rangle 
\end{pmatrix} = 
\begin{pmatrix}
\phi_{\sigma} \ ,&  \Delta\psi_{\sigma} 
\\[2pt]
 \Delta\bar{\psi}_{\sigma} \ ,&\Delta\bar{\phi}_{\sigma}
\end{pmatrix}
$$
\end{minipage}%\hspace{55mm}
 \\ = \quad
\begin{minipage}{40mm}
\includegraphics[width=40mm]{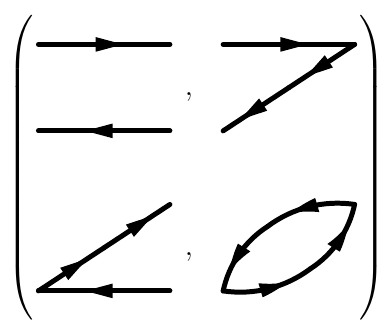} 
\end{minipage}
\end{multline}\\
where the angular brackets stand for the summation over the fermionic degrees of freedom of the one-particle propagators.

%\section{Critical behavior in the ordered state}

%\section{Bispinor of the electron-hole propagator and non-conserving terms}

\begin{figure}
\includegraphics[width=80mm]{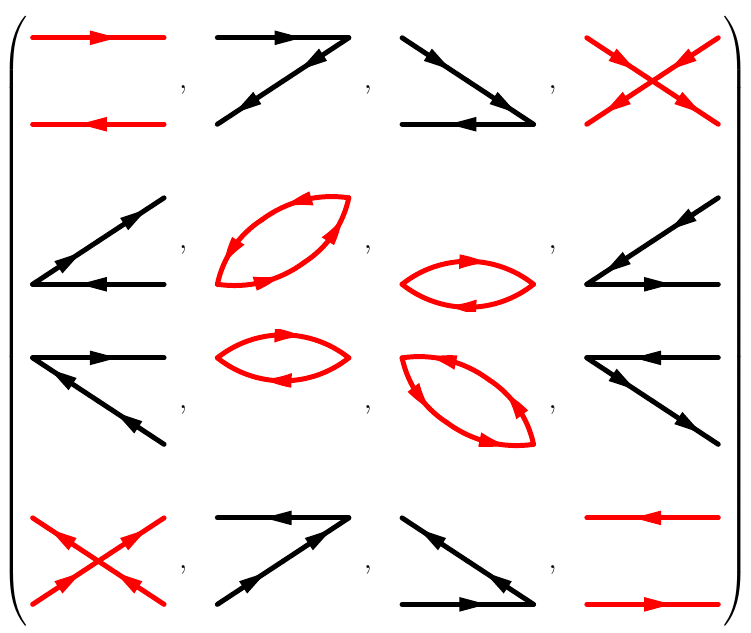}
\caption{Bispinor of the full spin-dependent electron-hole propagator in the ordered phase.  The red diagrams conserve the number of electrons and holes and guarantee that the ordered phase continuously matches the high-temperature solution at the critical transition point.   \label{fig:FEynm2P-full}}
\end{figure}

The spinor from Eq.~\eqref{eq:phi-Deltapsi} is not the complete or the only possible representation of the electron-hole propagation. In particular, when spins up and down are not equivalent and the electron-hole bubbles depend on the spin index.  The singlet and triplet terms are not identical. It means that the propagator of a pair of an electron with spin down and a hole with spin up is not equivalent to that of an electron with spin down and a hole with spin up.
To complete the representation of both electron-hole pairs with different spins, we have to use a bispinor representing their propagation \cite{Janis:2001ab}. Its diagrammatic representation is plotted in Fig.~\ref{fig:FEynm2P-full}. 

The contributions of the electron-hole propagation to the thermodynamic potential are thereby doubled since we added electron and hole representations with different spins. The electron-hole singlet spinors are positioned along the main diagonal. The off-diagonal spinors represent the triplet contributions with bubbles $\Delta\psi^{t}$ and $\Delta\phi^{t}$. The bispinor in Fig.~\ref{fig:FEynm2P-full} contains the electron-hole propagation with all formally accessible spin-flip processes. The problem with a mismatch of the critical point at which the anomalous contributions vanish with the critical point of the high-temperature susceptibility remains. The mismatch is because not all electron-hole propagators are fully conserving in thermodynamically closed systems.   

To match the ordered phase with the disordered one continuously at the transition point, we need to suppress the mixed diagrams $\Delta \bar{\psi}_{\sigma}$ and $\Delta\psi_{\sigma}$. They are specific in that they do not guarantee the conservation of the number of electrons and holes in the thermodynamically closed systems.  If an electron with spin $\sigma$ undergoes a spin-flip into a state with the opposite spin $\bar{\sigma}$, the latter spin state must be empty, be occupied by a hole of spin $\bar{\sigma}$. This hole with spin $\bar{\sigma}$ must be part of the closed system. The electron spin flip $\sigma\to\bar{\sigma}$ must simultaneously be accompanied by a hole spin flip $\bar{\sigma}\to\sigma$. Suppose the accompanying transition of the hole state is not explicitly part of the electron-hole propagator (mixed normal-anomalous propagators) then it is not guaranteed that the hole, needed for the electron spin flip, is part of the investigated closed system. The anomalous propagators must hence be considered only in pairs in closed systems, the red diagrams in Fig.~\ref{fig:FEynm2P-full}.

\begin{figure}
\includegraphics[width=45mm]{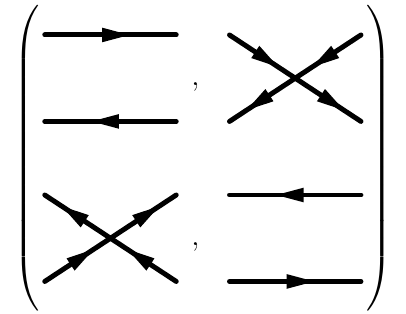}
\caption{Spinor of the electron-hole propagator in the ordered phase for spin-symmetric systems. \label{fig:FEynm2P-corr}}
\end{figure}

The full bispinor electron-hole propagators are needed only in the spin-biased systems where the electron-hole propagators depend on the spin index. The triplet and singlet electron-hole propagators are the same in the spin-symmetric systems, and we can return to the standard spinor representation as we did in the paper. The corresponding electron-hole spinor propagator used in the FLEX approximation, Eq.~\eqref{eq:Phi-matrix}, is plotted in Fig.~\ref{fig:FEynm2P-corr}.  The linear term must, however, be singled out in this representation of the thermodynamic potential.

%\end{widetext}
 
%\bibliographystyle{apsrev}
%\bibliography{../../../../../BibTeX/parquets_MB,../../../../../BibTeX/superconductivity,../../../../../BibTeX/Impurity_solver-1}

%\twocolumngrid

\end{document}